\documentclass[review,12pt]{elsarticle}
\usepackage{amssymb}

\usepackage{bm}

\journal{Physica E: Low-dimensional Systems and Nanostructures}

\begin{document}

\begin{frontmatter}

\title{Analysis of the Collective Behavior of a 10 by 10 Array\\
of Fe$_3$O$_4$ Dots in a Large Micromagnetic Simulation}

\author{Christine C. Dantas}
\ead{christineccd@iae.cta.br}
\address{Divis\~ao de Materiais (AMR), Instituto de Aeron\'autica e Espa\c co (IAE), Departamento de Ci\^encia e Tecnologia Aeroespacial (DCTA), Brazil, FAX/Tel: 55 12 3941-2333}


\begin{abstract}
We report a full (3D) micromagnetic simulation of a set of $100$ ferrite (Fe$_3$O$_4$) cylindrical dots, arranged in a 10 by 10 square (planar) array of side $3.27$ $\mu$m, excited by an external in-plane magnetic field. The resulting power spectrum of magnetic excitations and the dynamical magnetization field at the resulting resonance modes were investigated. The absorption spectrum deviates considerably from that of a single particle reference simulation, presenting a mode-shifting and splitting effect. We found an inversion symmetry through the center of the array, in the sense that each particle and its inversion counterpart share approximately the same magnetization mode behavior. Magnonic designs aiming at synchronous or coherent tunings of spin-wave excitations at given spatially separated points within a regular square array may benefit from the new effects here described.
\end{abstract}


\end{frontmatter}

\section{Introduction}

For several decades, there has been a great interest in the study of the collective spin excitations in magnetically ordered media, and recently theoretical and experimental investigations have been thoroughly conducted \cite{Hil2002,Gub2005a,Sel2006,Kro2007,Neu2010}. These investigations established the prospect of controlling SWs in magnonic crystals (similarly to the control of light in photonic crystals \cite{Joa1995}), motivating a whole new field, currently being referred to as magnonics (c.f. \cite{Kru2010} and references therein).

In magnonic crystals, dipolar (magnetostatic) interactions have an important physical role, since they couple excitation modes of individual, closely-spaced particles, affecting both the static and dynamic behavior of the magnetization \cite{Mat1997,Jor2000,Bai2006}. This effect results in the formation of collective modes in the form of Bloch waves \cite{Gub2005b,Gio2007,Wan2009,Ziv2011}, leading to allowed and forbidden magnonic states at given frequencies, or band gaps \cite{Vys2005,Gub2010}.  These and other particular characteristics stimulated new research directions in the study of ``spin-waves'' \cite{Gur1996} (hereon SWs), given the possibility of designing filters and waveguides for microwave nanotechnology applications \cite{Bar2007,Khi2010}. 

However, only recently experimental and numerical investigations on the modification of normal modes of magnetic excitations in periodically arranged nanomagnets  have been carried out in some detail \cite{Jung2002a,Gub2006,Tac2010,Kru2005a,Kru2005b,Kru2007,Kea2008,Kru2010b}, given the need for advanced experimental and computational capabilities.
The general theoretical formulation of magnetic phenomena at scales of $\sim 10^{-6}-10^{-9}$ m is based on the Landau-Lifshitz-Gilbert (LLG) equation \cite{LL1935,Gil1955,Bro1963} for the magnetization dynamics. Note that the LLG equation only be solved analytically for special cases \cite{dAq2004}, hence computational micromagnetics with increasingly detailed simulations have been an important aid at understanding micromagnetic phenomena \cite{Fid2000,Aha2001}.

In a series of papers by Kruglyak et al. \cite{Kru2005a,Kru2005b,Kru2007,Kea2008}, particular attention was given to the investigation of the magnetization dynamics of square arrays of submicron elements of different sizes under a range of bias fields. These investigations involved the use of time-resolved scanning Kerr microscopy to probe the magnetic response of nanoelements, along with micromagnetic simulations to aid the analysis of the resulting spectra. These works have generally shown that the position of the mode frequencies as a function of the element size as well as their relative absorption amplitudes present a complex dependence and follows a nonmonotonic behavior. It was also observed that the position of mode frequencies did not follow the prediction of the macrospin model for an isolated element. It was inferred that a nonuniform distribution of the demagnetizing field could be responsible for nonuniform precession within the elements of the array, adding to the complex dependence the role of exchange interaction. It has also been noted that, as the size of the element in the array is decreased, the edge regions of a given element present increasingly dominant modes confined by the demagnetizing field in relation to uniform modes. In addition, in studies where the orientation of the external magnetic field was rotated in the element plane of the array, the effective magnetic field inside a given element also presented an ``extrinsic'' anisotropic contribution due to the stray field from nearby elements, as contrasted to an ``intrinsic'' anisotropy occurring in an isolated element. Furthermore, a {\it dynamical} configurational anisotropy was necessary to qualitatively explain the data. An additional important feature, specially observed in micromagnetic simulations of arrays of nanoelements, is the splitting of precessional modes, a feature experimentally verified recently as detectable collective spin wave modes extended throughout the array \cite{Kru2010b}.

Here we report a $100$-particles micromagnetic simulation -- to our knowledge, the largest detailed micromagnetic calculation ever performed at the given scale and number of particles, with a careful observation of the accuracy requirements (see accuracy details in Section 2.3 of Ref. \cite{Dan2010}, which were also adopted here).  This work is part of an ongoing project \cite{Dan2008,Dan2010} motivated by the theoretical investigation and design of new nanostructured magnetic configurations with interesting SWs or magnonic band gap behavior, suitable for different applications in the microwave frequency range.  We show that the collective magnetization behavior is constrained by an inversion symmetry through the center of the array. In particular, this opens interesting possibilities for applications that require spatially coordinated patterns.

\section{Methods}

The simulations were performed by using the freely available integrator {\sl OOMMF} (Object Oriented Micromagnetic Framework)\cite{OOMMF}, which was installed and executed on a 3 GHz Intel Pentium Desktop PC, running Kurumin Linux. The present  $100$-particles simulation turned out to be a computationally demanding one (taking about $4$ months for completion, considering interruptions), and no other variation of the parameters were attempted at this time.  We applied the same methodology described in our previous works \cite{Dan2008,Dan2010}, based in the procedure given by Jung et al. \cite{Jung2002b}. 

An incident in-plane magnetic field was applied uniformly to the ferrite particles, composed by a static ({\it dc}) magnetic field ($B_{dc} \equiv \mu_0H_{dc}$) of $100$ mT in the $y$ direction, and a varying ({\it ac}) magnetic field ($B_{ac} \equiv \mu_0 H_{ac}$) of small amplitude ($1$ mT) in the $x$ direction, with the functional form: $B_{ac} = (1-e^{-\lambda t}) B_{ac,0} \cos (\omega t)$. We varied the frequency ($f = \omega / (2 \pi)$) from $2$ to $9.8$ GHz, in steps of $0.2$ GHz, resulting in $40$ different OOMMF frequency runs. The time domain of the applied $B_{ac}$ field was discretized in intervals of $0.005$ ns. 

OOMMF performed the numerical integration of the LLG equation leading to the evolution of the magnetization field of $100$ ferrite particles regularly distributed in a square, $10$ by $10$ array of side $3.27$ $\mu$m. The particles were identical cylindrical dots (each with a diameter $d = 0.3$ $\mu$m and thickness $\delta = 85$ nm). We adopted a small interparticle (edge-to-edge) spacing ($s = 0.03$ $\mu$m $\ll d$). The simulation was stopped at $5$ ns, giving $1000$ outputs for each frequency run. A reference simulation of a single-particle with a diameter $d = 0.3$ $\mu$m  was also conducted, with the same global parameters of the $100$-particles simulation. Table \ref{Tab1} lists the parameters used to set up the OOMMF integrator in both cases. 

The power spectra of magnetic excitations were obtained by excluding data from the first $2$ ns of the averaged magnetization vector in the $x$ direction, $\langle \vec{M} \rangle _{\rm x}(t\leq 2~ {\rm ns})$, and taking the Fourier transform of the remaining time domain data, $\langle \vec{M} \rangle _{\rm x}(2 < t \leq 5 ~{\rm ns})$. The maximum Fourier peak at each frequency provided the magnitude of absorption at the given frequency. A spline fit to the absorption data was performed in order to facilitate the comparison of the overall behavior of the curves, but individual data points were maintained.

\section{Results}

\begin{figure}[htbp]
\centering
\includegraphics[scale=0.45]{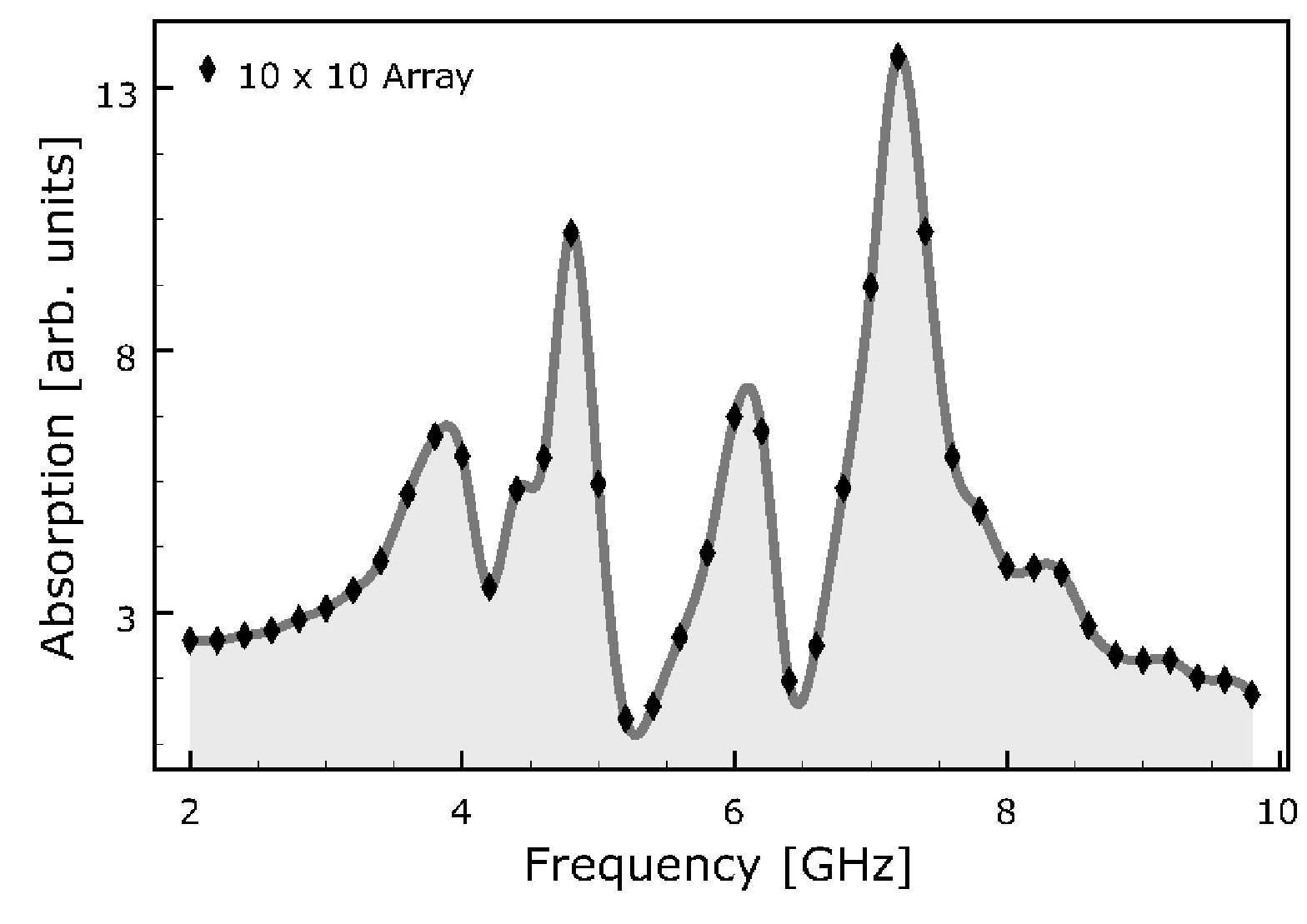}
\includegraphics[scale=0.45]{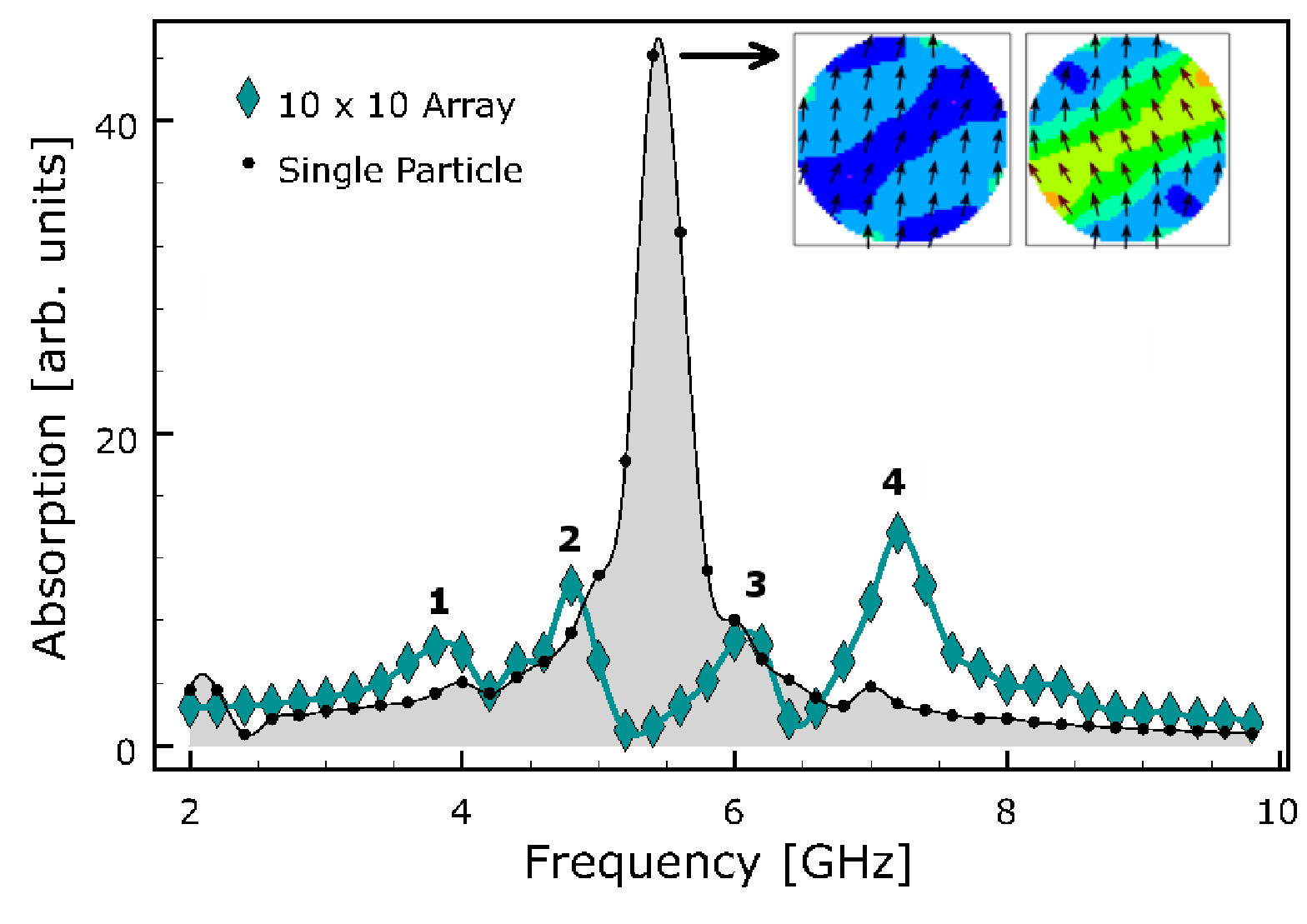}
\caption{\label{fig1}}
\end{figure}

The power spectra of magnetic excitations for the $100$-particles simulation and single-particle simulation are shown in Fig. \ref{fig1}. The resonance peak in the power spectrum of the single-particle simulation is clearly split into four distinct peaks of lower amplitude in the $100$-particles simulation, with two peaks at a lower and the other two at a higher frequency with respect to the reference fundamental mode, which is located at the gap between resonances 2 and 3 of the $100$-particles simulation. 

\begin{center}
\begin{figure}[htbp]
\centering
\includegraphics[scale=0.55]{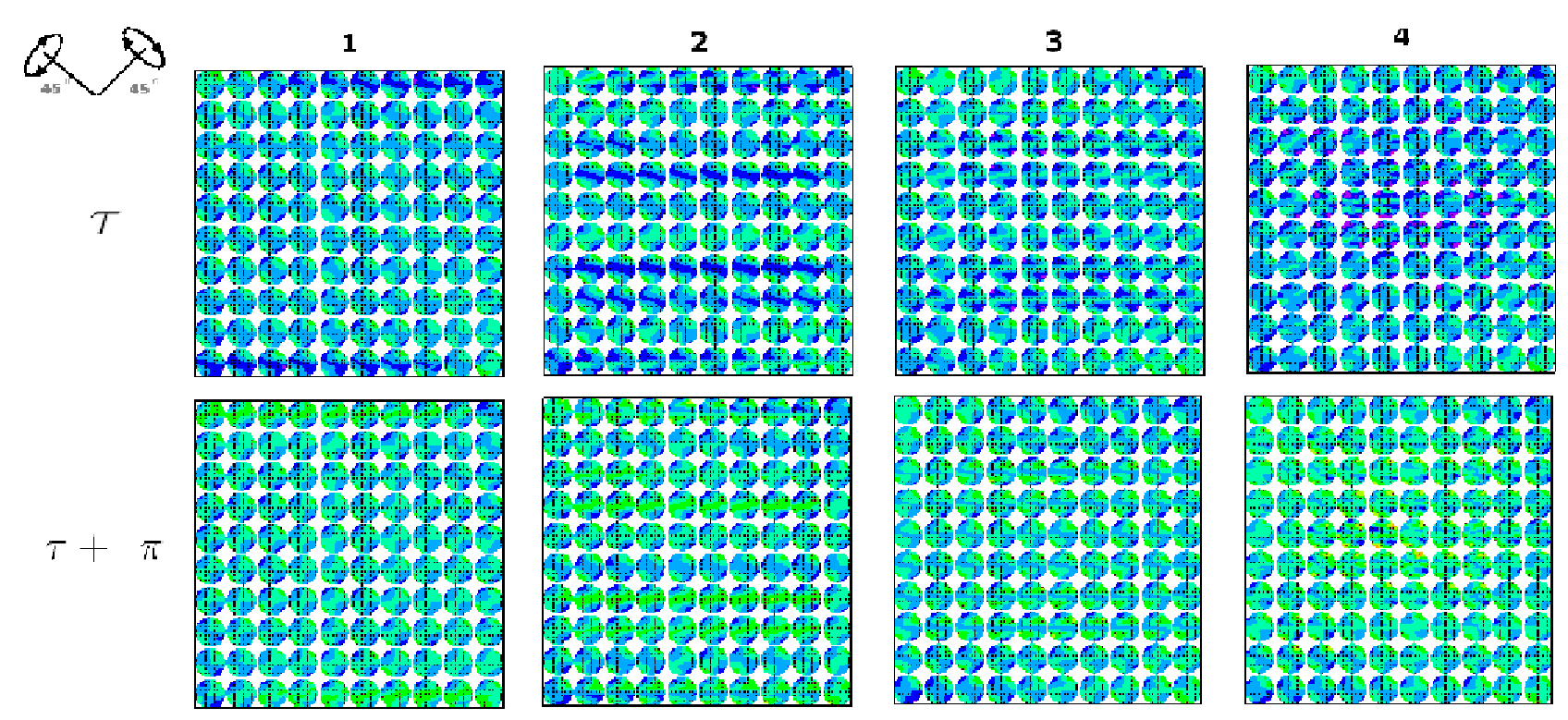}
\caption{\label{fig2}}
\end{figure}
\end{center}

We analysed the nature of the modes of interest by an inspection of the time-dependent magnetization vector field configuration. Bitmaps or ``snapshots" of the corresponding simulations were generated, selected at two points of the {\it ac} field cycle, namely, at the highest ($\tau$) and lowest ($\tau + \pi$) representative amplitudes of the equilibrium magnetization oscillation. We subsampled the $x$-component of the magnetization field in order to show $9$ vectors per cell element, and different pixel tonalities correspond to different values of $M_x$. In Fig. \ref{fig1} (inset), the magnetization vector field of the one-particle simulation around its resonance peak is shown. This should be contrasted with those of Fig. \ref{fig2} (snapshots of the $100$-particles simulation at the previously identified four peaks of interest). 

\begin{center}
\begin{figure}[htbp]
\centering
\includegraphics[scale=0.55]{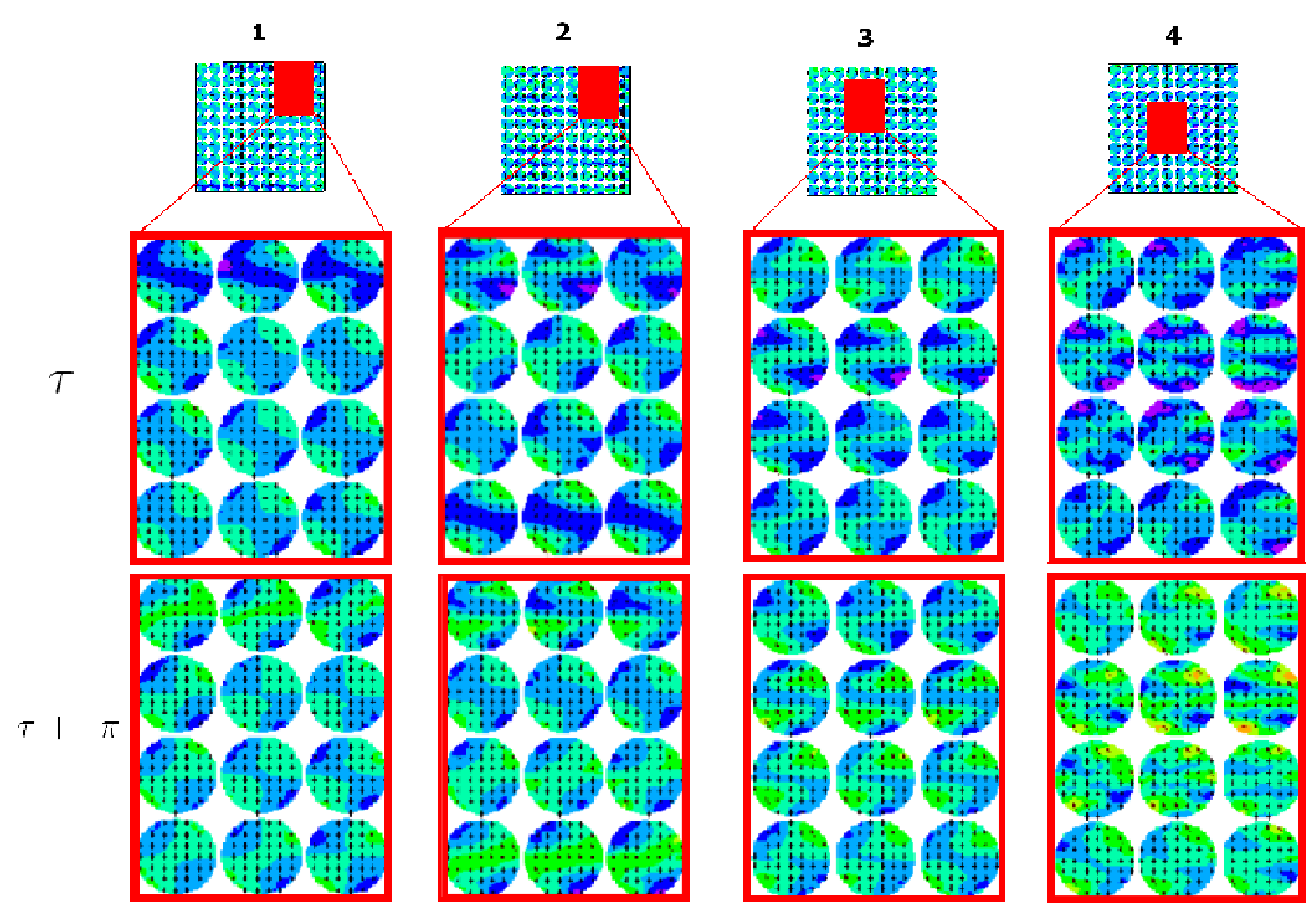}\begin{center}(a)\end{center}\vspace{1cm}
\includegraphics[scale=0.45]{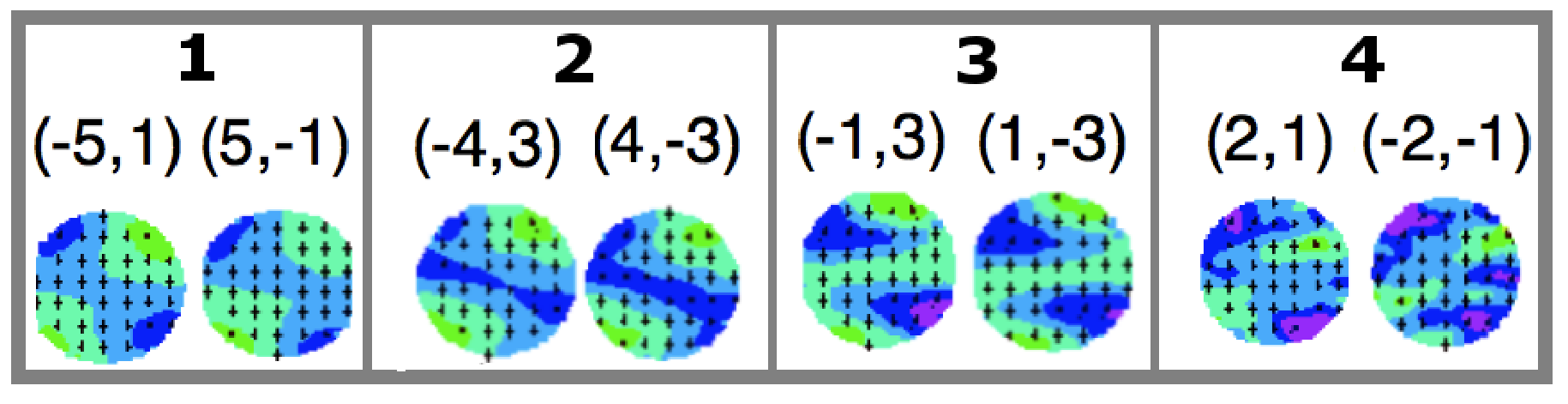}\begin{center}(b)\end{center}
\caption{\label{fig3}}
\end{figure}
\end{center}

Fig. \ref{fig3}(a) shows some zoom-out regions of the $100$-particles simulation in order to allow for the identification of several types of magnetization mode behavior, which will be discussed in more detail in the next section. On the other hand, from a visual inspection of the $10$ by $10$ snapshots, it is possible to identify an inversion symmetry through the center of the array in such a way that each particle and its inversion counterpart share approximately the same mode behavior in the array. In other words, by setting a Cartesian grid on the array, where the origin of the coordinates is the center of the array, and where each element is centered at coordinates $(i,j)$, an inversion transformation $(i,j) \rightarrow (-i,-j)$ leaves the magnetization configuration of the array approximately invariant.  Some examples are shown in Fig. \ref{fig3}(b). This global symmetry should arise from dipolar couplings, but the exact formulation is yet not clearly understood. This effect has already been pointed out in our previous work in the cases of $2$ by $2$ and $3$ by $3$ array simulations (c.f. Fig 8 of Ref. \cite{Dan2008} and discussions in Ref. \cite{Dan2010}), but it was not possible at that time to extrapolate whether the effect would persist in a larger array. 

In order to address in a more quantitative way the magnetization dynamics distribution in the array, we computed two simple estimators, which nevertheless establish the relevance of the visually noted effect. The first estimator is the modulus of the difference of average magnetization values at the points of the {\it ac} field cycle: $m _{(i,j)} \equiv \mid \langle M_x \rangle _{(i,j)}(\tau) - \langle M_x \rangle _{(i,j)}(\tau + \pi) \mid$; where $\langle M_x \rangle _{(i,j)}$ is the average magnetization of a given particle at grid coordinates $(i,j)$. The second estimator, $\sigma_{(i,j)}$, is the modulus of the difference of standard deviation values of particle magnetizations within the array, that is, the standard deviation is computed with reference to the average magnetization of the {\it whole} array. Fig. \ref{fig4} shows the resulting $m _{(i,j)}$ and $\sigma_{(i,j)}$ maps computed for the $4$ modes of interest. Notice that each map pixel is labelled by the  particle coordinate $(i,j)$. We discuss these maps in more detail in the next section.

\section{Discussion and Conclusion}

It is presently understood that particular features of the power spectra of magnetic excitations can be associated with the various types of nodal behavior of the time-dependent magnetization field (see, e.g., Refs. \cite{Gub2006,Jor2000,Jung2002a,Kru2005a,Kru2005b,Kru2007,Kea2008}).
A general ``spin-wave" behavior (SWB) for the excitations indicates that the magnetization vectors present small amplitude oscillations about a nonuniform static magnetization field. In addition, the following subclasses of excitations can be identified \cite{Jung2002a,Dan2010}: (i) {\sl``Quasi-uniform" behavior (QUB)}: the movement of each magnetization vector is similar to that of its neighbors, except for the regions around the edge of the particle; and (ii) {\sl ``Edge-like" behavior (EDB)}: the magnetization field in the central region of the particle is almost entirely static and aligned with the direction of the external {\it dc} field; the nonuniformly distributed magnetization vectors present small amplitude oscillations near the edges of the particle. In particular, these modes may be more affected by the dipolar couplings from nearby particles. 

In the present work, the nature of the reference absorption peak is basically due to QUB, whereas the $100$-particles simulation presents all types of behavior (as can be seen from an inspection of the zoom-out regions exemplified in Fig. \ref{fig3}(a)). In particular, peak $1$ at $3.8$ GHz is dominated by QUB, with a few EDB cases specially for particles at the borders of the array. Peak $2$ at $4.8$ GHz shows a mixture of QUB and EDB, with a few SWB in some particles at the top and bottom of the array. The number of SWB cases appears to increase in peak $3$ at $6.0$ GHz and is the dominant absorption mechanism of peak $4$ at $7.2$ GHz, specially for particles located towards the center of the array. This effect is approximately the same as that observed previously for $3$ by $3$ arrays (c.f. Fig. 9 of Ref. \cite{Dan2010}). 

\begin{figure}[htbp]
\includegraphics[scale=0.25]{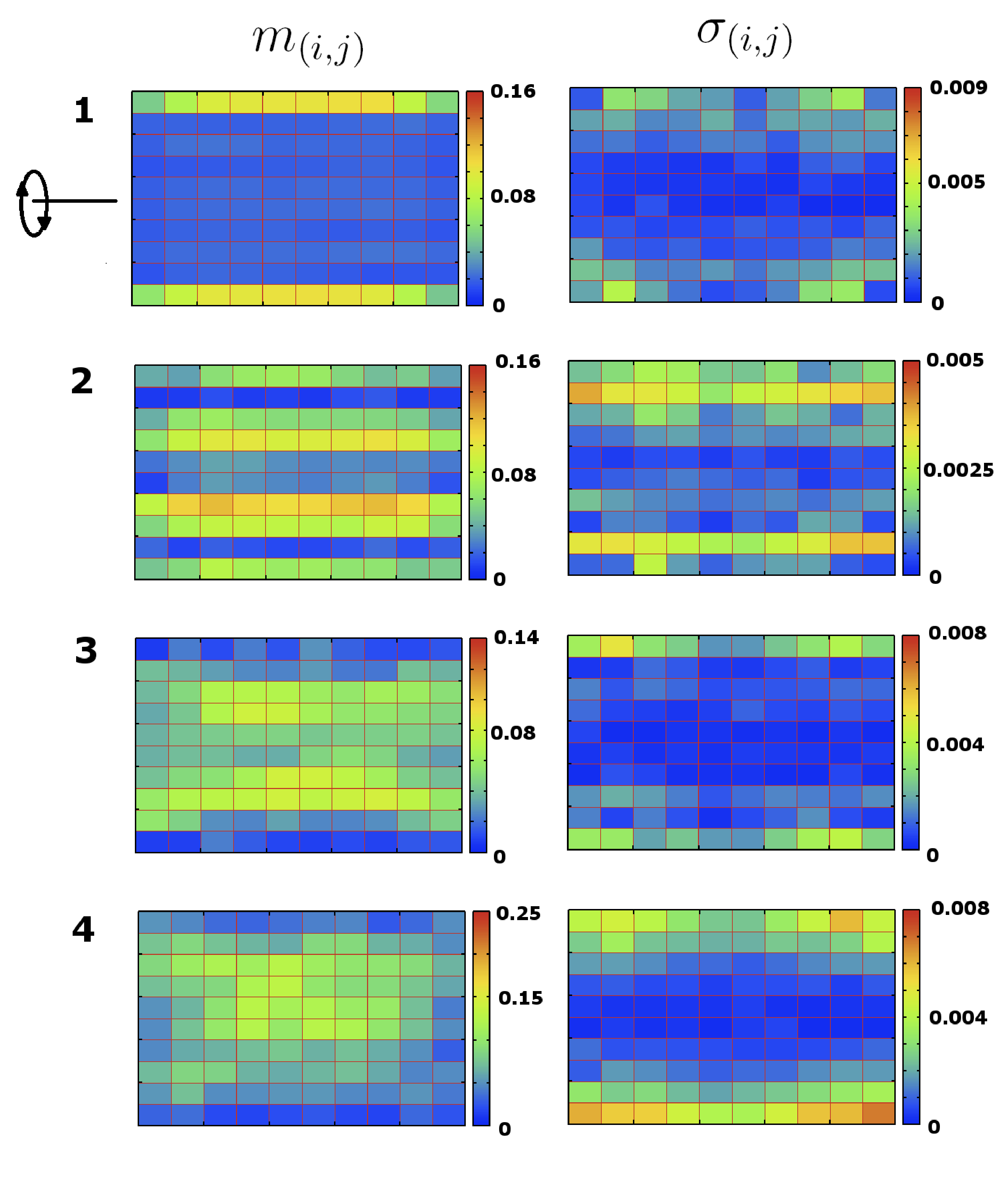}
\caption{\label{fig4}}
\end{figure}

In order to address in a more quantitative way the inversion symmetry in the array (c.f. Fig. \ref{fig2}), as pointed out in the previous section, we refer to the  $m _{(i,j)}$ and $\sigma_{(i,j)}$ estimators shown in Fig. \ref{fig4}.  It is clear that $m _{(i,j)}$ gives larger values for larger amplitudes of the average particle magnetization, but also the original inversion symmetry through the center of the array of the magnetization configuration should be translated approximately into a symmetry around a central horizontal axis in a $m _{(i,j)}$ map. It is clear that if the standard deviation distribution is similar at the two opposite amplitude points, then a map of $\sigma_{(i,j)}$ should be uniform. This estimator also gives larger values for particles whose behavior with respect to the whole array presents a substantial difference at the two points of the cycle, thus furnishing an overall measure of the degree of cycle ``coherence".

The expected central horizontal axis symmetry is indeed seen in Fig. \ref{fig4} (left column of panels), which corroborates our visual analysis. In peak 1,  $m _{(i,j)}$ is larger for particles localized in the top and bottom rows of the array. In peak 2, note the interesting regular, alternating magnetization excitations along specific rows of the array (also inferred from a visual inspection of the snapshots). Peaks 3 and 4 show higher values of $m _{(i,j)}$ in the central regions, in contrast to peak 1.  The $\sigma_{(i,j)}$ estimator results (Fig. \ref{fig4}, right column of panels) show that, excluding the top and bottom rows of the arrays in cases 1, 3 and 4, the array behavior is reasonably similar at the two extremes of the cycle. However, case 2 shows again a pattern, where the second and ninth rows (related to particles with very low amplitudes of the average particle magnetization, c.f. left column of panels) have standard deviations that significantly differ at the two extremes of the cycle.

Qualitatively, the results reported in the present work, specially the splitting of the resonance mode and its decreased relative amplitude, as well as the spatially nonuniform behavior of the elements in the array, are compatible with the behavior of the collective excitations reported in similar previous works \cite{Kru2005a,Kru2005b,Kru2007,Kea2008,Kru2010b}. Particularly, in Ref. \cite{Kru2010b}, for the first time the collective spin wave modes of the entire array has been experimentally detected. With the aid of a micromagnetic simulation of a $3 \times 3$ array, the authors observed how the row/column elements behaved out of phase with the rest of the elements of the array, qualitatively explaining the nature of the splitting of the precessional modes of the absorption spectrum. Our results show that this behavior is even more complex when considering a more extended array of $10 \times 10$ elements. Yet, as already mentioned, a symmetry pattern can be identified. Indeed, the inversion symmetry here noted is compatible with the experimentally observed ``tilt'' of the modes in regions of higher amplitude \cite{Kru2010b}, relatively to the horizontal and vertical axes, reported in that work. We believe that such a ``tilt'' would be observed in a larger simulation for that material.

In summary, our present results and previous indications allow us to infer that, for interparticle (edge-to-edge) separations at least of the order of $\sim 10$ to $20$\% of the particle's diameter, dipolar couplings in periodically arranged cylindrical nanomagnets will cause a global, coherent magnetization behavior across a square array distribution, with an inversion symmetry through the center of the array (for in-plane magnetic excitations).  The power spectrum shows a clear four-fold resonance feature, wherein the magnetization field distribution and dynamics show interesting patterns and trends. We hope that the effect here reported will stimulate the development of a theory that will generally describe and predict similar mode symmetries in periodic arrays. Our results may be of interest for magnonic device architectures, specially for applications dealing with pairwise SWs excitations of submicron ferrite particles at spatially separated points in a periodic square array.

\section{Acknowledgments} 

We thank the anonymous referee for useful comments and corrections. We acknowledge the support of Dr. Mirabel C. Rezende and FINEP/Brazil. 


\section{References}

\bibliographystyle{elsarticle-num}

\bibliography{Dantas_Ref_2011}

\begin{thebibliography}{10}
\expandafter\ifx\csname url\endcsname\relax
  \def\url#1{\texttt{#1}}\fi
\expandafter\ifx\csname urlprefix\endcsname\relax\def\urlprefix{URL }\fi
\expandafter\ifx\csname href\endcsname\relax
  \def\href#1#2{#2} \def\path#1{#1}\fi

\bibitem{Hil2002}
K.~Hillebrands, B. \&~Ounadjela, Spin Dynamics in Confined Magnetic Structures
  I, III, Berlin: Springer, 2002.

\bibitem{Gub2005a}
G.~{Gubbiotti}, et~al., {Spin dynamics in thin nanometric elliptical Permalloy
  dots: A Brillouin light scattering investigation as a function of dot
  eccentricity}, Phys. Rev. B 72~(18) (2005) 184419--184427.

\bibitem{Sel2006}
R.~Sellmyer, D. \&~Skomski, Advanced Magnetic Nanostructures, New York:
  Springer, 2006.

\bibitem{Kro2007}
S.~Kronmuller, H. \&~Parkin, Handbook of Magnetism and Advanced Magnetic
  Materials vol 3, Chichester: Wiley, 2007.

\bibitem{Neu2010}
S.~{Neusser}, et~al., {Anisotropic Propagation and Damping of Spin Waves in a
  Nanopatterned Antidot Lattice}, Physical Review Letters 105~(6) (2010)
  067208--067212.

\bibitem{Joa1995}
J.~D. {Joannopoulos}, R.~D. {Meade}, J.~N. {Winn}, Photonic Crystals: Molding
  the Flow of Light, Princeton: Princeton Univ. Press, 1995.

\bibitem{Kru2010}
V.~V. {Kruglyak}, S.~O. {Demokritov}, D.~{Grundler}, {Magnonics}, Journal of
  Physics D Applied Physics 43~(26) (2010) 264001--264015.

\bibitem{Mat1997}
C.~{Mathieu}, et~al., {Anisotropic magnetic coupling of permalloy micron dots
  forming a square lattice}, Applied Physics Letters 70 (1997) 2912--2914.

\bibitem{Jor2000}
J.~{Jorzick}, et~al., {Quantized spin wave modes in micron size magnetic
  disks}, Journal of Applied Physics 87 (2000) 5082--5084.

\bibitem{Bai2006}
M.~{Bailleul}, R.~{H{\"o}llinger}, C.~{Fermon}, {Microwave spectrum of square
  Permalloy dots: Quasisaturated state}, Phys. Rev. B 73~(10) (2006)
  104424--104438.

\bibitem{Gub2005b}
G.~Gubbiotti, et~al., Magnetostatic interaction in arrays of nanometric
  permalloy wires: A magneto-optic kerr effect and a brillouin light scattering
  study, Phys. Rev. B 72~(22) (2005) 224413--224420.

\bibitem{Gio2007}
L.~Giovannini, F.~Montoncello, F.~Nizzoli, Effect of interdot coupling on
  spin-wave modes in nanoparticle arrays, Phys. Rev. B 75~(2) (2007)
  024416--024423.

\bibitem{Wan2009}
Z.~K. {Wang}, et~al., {Observation of frequency band gaps in a one-dimensional
  nanostructured magnonic crystal}, Applied Physics Letters 94~(8) (2009)
  083112--083115.

\bibitem{Ziv2011}
R.~{Zivieri}, et~al., {Collective spin modes in chains of dipolarly interacting
  rectangular magnetic dots}, Phys. Rev. B 83~(5) (2011) 054431--054440.

\bibitem{Vys2005}
S.~L. {Vysotskii}, S.~A. {Nikitov}, Y.~A. {Filimonov}, {Magnetostatic spin
  waves in two-dimensional periodic structures (magnetophoton crystals)},
  Journal of Experimental and Theoretical Physics 101 (2005) 547--553.

\bibitem{Gub2010}
G.~{Gubbiotti}, et~al., {Brillouin light scattering studies of planar metallic
  magnonic crystals}, Journal of Physics D Applied Physics 43~(26) (2010)
  264003--264017.

\bibitem{Gur1996}
G.~A. Gurevich, A. G. \&~Melkov, Magnetization Oscillations and Waves, New
  York: Chemical Rubber Corp., 1996.

\bibitem{Bar2007}
K.~{Baberschke}, {Why are spin wave excitations all important in nanoscale
  magnetism?}, Physica Status Solidi (b) 245 (2007) 174--181.

\bibitem{Khi2010}
A.~{Khitun}, M.~{Bao}, K.~L. {Wang}, {Magnonic logic circuits}, Journal of
  Physics D Applied Physics 43~(26) (2010) 264005--264016.

\bibitem{Jung2002a}
S.~{Jung}, et~al., {Ferromagnetic resonance in periodic particle arrays}, Phys.
  Rev. B 66~(13) (2002) 132401--132405.

\bibitem{Gub2006}
G.~{Gubbiotti}, et~al., {Normal mode splitting in interacting arrays of
  cylindrical permalloy dots}, Journal of Applied Physics 99~(8) (2006)
  080000--080003.

\bibitem{Tac2010}
S.~{Tacchi}, et~al., {Anisotropic dynamical coupling for propagating collective
  modes in a two-dimensional magnonic crystal consisting of interacting squared
  nanodots}, Phys. Rev. B 82~(2) (2010) 024401--024409.

\bibitem{Kru2005a}
V.~V. {Kruglyak}, A.~{Barman}, R.~J. {Hicken}, J.~R. {Childress}, J.~A.
  {Katine}, {Precessional dynamics in microarrays of nanomagnets}, Journal of
  Applied Physics 97~(10) (2005) 10A706.

\bibitem{Kru2005b}
V.~V. {Kruglyak}, A.~{Barman}, R.~J. {Hicken}, J.~R. {Childress}, J.~A.
  {Katine}, {Picosecond magnetization dynamics in nanomagnets: Crossover to
  nonuniform precession}, Phys. Rev. B 71~(22) (2005) 220409(R).

\bibitem{Kru2007}
V.~V. {Kruglyak}, P.~S. {Keatley}, R.~J. {Hicken}, J.~R. {Childress}, J.~A.
  {Katine}, {Dynamic configurational anisotropy in nanomagnets}, Phys. Rev. B
  75~(2) (2007) 024407.

\bibitem{Kea2008}
P.~S. {Keatley}, V.~V. {Kruglyak}, A.~{Neudert}, E.~A. {Galaktionov}, R.~J.
  {Hicken}, J.~R. {Childress}, J.~A. {Katine}, {Time-resolved investigation of
  magnetization dynamics of arrays of nonellipsoidal nanomagnets with
  nonuniform ground states}, Phys. Rev. B 78~(21) (2008) 214412.

\bibitem{Kru2010b}
V.~V. {Kruglyak}, P.~S. {Keatley}, A.~{Neudert}, R.~J. {Hicken}, J.~R.
  {Childress}, J.~A. {Katine}, {Imaging Collective Magnonic Modes in 2D Arrays
  of Magnetic Nanoelements}, Physical Review Letters 104~(2) (2010) 027201.

\bibitem{LL1935}
E.~Landau, L. \&~Lifshitz, On the theory of the dispersion of magnetic
  permeability in ferromagnetic bodies, Physik. Z. Sowjetunion 8 (1935)
  153--160.

\bibitem{Gil1955}
T.~L. Gilbert, A lagrangian formulation of the gyromagnetic equation of
  magnetization field, Phys. Rev. 100 (1955) 1243--1253.

\bibitem{Bro1963}
W.~F. Brown, Micromagnetics, Interscience Publishers, 1963.

\bibitem{dAq2004}
M.~d'Aquino, Nonlinear magnetization dynamics in thin-films and nanoparticles,
  Doctorate Thesis in Electrical Engineering, Universit\`a Degli Studi di
  Napoli, It\'alia, 2004.

\bibitem{Fid2000}
J.~{Fidler}, T.~{Schrefl}, {TOPICAL REVIEW: Micromagnetic modelling - the
  current state of the art}, Journal of Physics D Applied Physics 33 (2000)
  135--156.

\bibitem{Aha2001}
A.~Aharoni, Micromagnetics: past, present and future, Physica B 306 (2001)
  1--9.

\bibitem{Dan2010}
C.~C. {Dantas}, A.~M. {Gama}, {Micromagnetic simulations of spinel ferrite
  particles}, Journal of Magnetism and Magnetic Materials 322 (2010)
  2824--2833.

\bibitem{Dan2008}
C.~C. {Dantas}, L.~A. {de Andrade}, {Micromagnetic simulations of small arrays
  of submicron ferromagnetic particles}, Physical Review B 78 (2008)
  024441--024449.

\bibitem{OOMMF}
M.~Donahue, D.~Porter, \href{http://math.nist.gov/oommf/}{{OMMF User's Guide,
  Version 1.0, {\tt http://math.nist.gov/oommf/}}}, Interagency Report NISTIR
  6376, National Institute of Standards and Technology, Gaithersburg, MD.
\newline\urlprefix\url{http://math.nist.gov/oommf/}

\bibitem{Jung2002b}
S.~{Jung}, J.~B. {Ketterson}, V.~{Chandrasekhar}, {Micromagnetic calculations
  of ferromagnetic resonance in submicron ferromagnetic particles}, Phys. Rev.
  B 66~(13) (2002) 132405--132409.

\end{thebibliography}

\clearpage

\begin{table}
\caption{\label{Tab1} {Main OOMMF parameters, fixed for all simulations.}}
\begin{tabular}{lll} \hline \hline
Simulation Parameter/Option    & \hspace{1cm} & Parameter \\ \hline \hline
Saturation magnetization [A/m] &  & $5.0 \times 10^5$\\
Exchange stiffness [J/m]       &  & $1.2 \times 10^{-11}$ \\
Anisotropy constant [J/m$^3$]  &  & $-1.10\times 10^{4}$  \\
Anisotropy Type                &  & cubic \\
First Anisotropy Direction (x,y,z)      &  & (1 1 1)   \\
Second Anisotropy Direction (x,y,z)     &  & (1 0 0)    \\
Damping constant               &  & $0.005$\\
Gyromagnetic ratio [m/(A.s)]   &  & $2.21 \times 10^5$\\
Particle thickness [nm]        &  & $85.0$ \\
Cell size [nm]                 &  & $5.0$ \\
Demagnetization algorithm type &  & const. in each cell\\ \hline
\end{tabular}
\end{table}

\clearpage

{\it Fig. 1}: {\it Top panel:} Power spectrum of magnetic excitations of the $100$-particles simulation. {\it Bottom panel:} Comparison of the former spectrum (diamond symbols) with that of the single-particle reference simulation (dot symbols). Four distinct peaks in the spectrum of the $100$-particles simulation are identified. ``Snapshots'' of the magnetization vector field of the one-particle simulation around the resonance peak are indicated by the arrow. Snapshot to the left refers to the highest amplitude of the oscillation and the snapshot to the right, to the lowest one. 

{\it Fig. 2}: ``Snapshots'' of the magnetization vector field of the $100$-particles ($10$ by $10$ particle array) simulation around the four peaks identified in the previous figure. For each peak, the upper snapshot shows the magnetization state of the array at the highest amplitude of the oscillation ($\tau$) whereas the snapshot immediately below, at the lowest amplitude ($\tau + \pi$). Symmetry axes are shown schematically at the top left of the figure.

{\it Fig. 3}: (a) Amplified particular selections ($3$ by $4$ sub-arrays given by the solid rectangular shown inside the $10$ by $10$ array snapshots), at the two points of the cycle. (b) Illustration of the symmetric mode behavior: one example of particle pairs is taken from each of the four peaks at $\tau$. Grid coordinates are indicated above the particles.

{\it Fig. 4}: Estimator maps ($m _{(i,j)}$ and $\sigma_{(i,j)}$ distributions) for the arrays, as explained in the main text. New symmetry axis is indicated.

\clearpage

\clearpage

\clearpage

\clearpage

\end{document}